\documentclass[useAMS,usenatbibi,usegraphicx]{mn2e}
\usepackage{graphics}
\usepackage{psfig}
\usepackage{epsfig}

\newcommand\beq{\begin{equation}}
\newcommand\eeq{\end{equation}}

\def\msun{\,{\rm M_\odot}}

\def\gsim{ \lower .75ex \hbox{$\sim$} \llap{\raise .27ex \hbox{$>$}} }
\def\lsim{ \lower .75ex\hbox{$\sim$} \llap{\raise .27ex \hbox{$<$}} }

\title[Halo millisecond pulsars]
{Halo millisecond pulsars ejected by intermediate mass black holes in globular clusters}

\author[A. Sesana et al.]{A. Sesana$^{1}$, N. Sartore$^{2}$, B. Devecchi$^{3}$, A. Possenti$^{4}$\\
$^{1}$Albert Einstein Institut, Am M\"uhlenberg 1, Golm, D-14476, Germany.\\
$^{2}$INAF - Istituto di Fisica Spaziale e Fisica Cosmica, via E. Bassini 15, I-20133 Milano, Italy.\\
$^{3}$Leiden Observatory, Niels Bohrweg 2, NL-2333 CA, Leiden, The Netherlands.\\
$^{4}$INAF - Osservatorio Astronomico di Cagliari, Loc. Poggio dei Pini, Strada 54 - 09012 Capoterra (CA), Italy.}

\begin{document}

\date{Received ---}

\maketitle

\begin{abstract}
Intermediate mass black holes (IMBHs) are among the most elusive objects in
contemporary astrophysics. Both theoretical and observational evidence of their
existence is subject of debate. Conversely, both theory and observations confirm
the presence of a large population of millisecond pulsars (MSPs) with low mass 
companions residing in globular cluster (GC) centers. If IMBHs are common in
GC centers as well, then dynamical interactions will inevitably break up many
of these binaries, causing the ejection of several 
fast MSPs in the Galactic halo. Such
population of fast halo MSPs, hard to produce with 'standard' MSP generation
mechanisms, would provide a strong, albeit indirect, evidence
of the presence of a substantial population of IMBHs in GCs. In this paper we
study in detail the dynamical formation and evolution of such fast MSPs 
population, highlighting the relevant observational properties and assessing
detection prospects with forthcoming radio surveys.  
\end{abstract}


\section{Introduction}
Black holes (BHs), with masses ranging from tens to billion solar
masses, are among the most exciting objects in the astrophysical
world. Although there is nowadays plenty of evidence for the existence
of stellar mass BHs ($M\sim 10\msun$) and massive BHs (MBHs, $M>
10^6\msun$), there is no convincing observational proof for the
existence of BHs in the $100\msun$-$10^5\msun$ range, the so called
intermediate mass BHs (IMBHs).  IMBHs could form as a result
  of dynamical instabilities in dense stellar clusters. Despite the
  large number of works devoted to this problem, any firm conclusion
  is still lacking.
Runaway collisions between massive
stars after their sinking into the cluster center provides a channel to
form a very massive ($\sim10^3\msun$) star in the center. This
  object would be expected to collapse into an intermediate mass
black hole ( Miller \& Hamilton 2002; Portegies Zwart \& McMillan
2002; Portegies Zwart et al. 2004) retaining most of the
  progenitor mass. More detailed simulations including stellar
evolution and feedback from stellar winds at solar metallicities, show a
systematic inefficiency in growing the central object to masses larger
than $\sim 100\msun$ (e.g., Glebbeek et al. 2009). IMBHs could
  then form only in those initially metal poor systems for which
  stellar mass loss via stellar wind would be strongly reduced.

If theoretical results are unclear, observational evidences are at
best elusive.  Globular clusters (GCs), among the densest stellar
systems known in galaxies, have become prime sites for IMBH
searches. The presence of an IMBH would affect the dynamics of the
cluster in several ways, and in the last decade many signatures have
been proposed. The presence of an IMBH would 'heat' the stellar
density profile, creating a large core (Baumgardt et al. 2005; Trenti
2008; Umbreit et al. 2009). Therefore, IMBHs are likely to reside
in those cluster showing a large $r_c/r_h$ ratio, where $r_c$ and
$r_h$ are the core and the half light radii,
respectively. Observations suggest (Baumgardt et al. 2005) that this
may be the case for $\sim30\%$ of the Milky Way (MW) GCs. 
Another clear fingerprint is the Keplerian rising of the velocity
dispersion of the stellar distribution inside the sphere of influence
of the IMBH. However, for typical IMBH masses, such signature would
appear on sub-arcsecond scales, at the limit of current optical
facilities. Ibata et al. (2009) report the detection of a Keplerian
cusp in M54, consistent with a central $10^4\msun$ IMBH (although
radial anisotropy in the stellar distribution may explain the
observations as well). Several other line-of-sight velocity studies
were undertaken by Baumgardt et al. (2003a), van den Bosch et
al. (2006), and Chakrabarty (2006) on M15, by Gebhardt et al. (2002),
Baumgardt et al.  (2003b), Gebhardt et al. (2005) on G1, by Noyola et
al. (2008), Sollima et al. (2009), van der Marel \& Anderson (2010) on
Omega Centauri, resulting for now in no undisputed definitive
detection. Globular clusters are old systems with relaxation timescales
shorter than their lifetimes. The most massive stars should then have had
enough time to segregate into the center via dynamical
friction. This mass segregation should be partially suppressed by
the presence of an IMBH. Several studies on the radial dependence
of the stellar mass function in different cluster might be consistent
with the presence of an IMBH (see Beccari et al. 2010 for M10,
Pasquato et al. 2009 for NGC 2298, Umbreit et al.  2009 for
NGC5694). However, the presence of primordial binaries can mimic the
same effect of an IMBH, and it is not clear to what extent the results
depend on the chosen initial conditions. 
Lastly, an IMBH would accrete ambient gas, and such accretion activity
may be observable in X-ray and/or radio.  Measurements of radio
luminosity of few Galactic GCs put stringent upper limits on IMBH
masses (Maccarone \& Servillat 2008). Similarly, Cseh et al. (2010)
impose a limit of 1500$\msun$ to any putative massive object in the
center of NGC 6388, by using combined X-ray and radio observations.
We notice, however, that these limits are affected by a number of
parameters (like the accretion efficiency, the gas content in GCs,
etc.) that are not constrained themselves.

Millisecond pulsars (MSPs) are usually associated with dense stellar
environment, as their formation channel requires a recycling phase
within a binary (Camilo \& Rasio 2006, Lorimer 2008): almost two thirds
(140 out of 220) of the known MSPs are observed in GCs, most of them
in binary systems residing in the GC cores (Lorimer 2008). The
simultaneous presence of an IMBH and of MSPs in binary systems in the
same GC opens interesting dynamical possibilities.  Dynamical
interactions with an IMBH are expected to break-up binaries ejecting
one of the components while leaving the other one bound to the IMBH
(Hills 1988, see Devecchi et al. 2007 in the context of IMBHs). One
consequence of the presence of IMBHs in GCs would then be the presence
of a halo population of fast MSPs, ejected by this tidal break-up
process. The detection of such population of fast halo MSPs would not
provide a direct evidence for the presence of an IMBH in any
particular cluster, but will prove the existence of a population of
IMBHs residing in GCs and capable of ejecting MSPs via
dynamical interactions. In this paper, we investigate the properties
of the expected fast halo MSP population for a range of formation
models, assessing observability with forthcoming radio
instruments, such as the Square Kilometre Array (SKA, Lazio 2009).

The paper is organized as follows. In Section 2 we introduce the main
ingredients of our models, namely the GC IMBH and MSP populations. The
binary-IMBH dynamical interaction is detailed in Section 3. In Section
4 we present the population of ejected MSPs highlighting their relevant
properties and we discuss observational prospects in Section 5. Our
main findings are summarized in Section 6.

\section{The actors on the stage}
\subsection{Globular cluster properties}
We take the Galactic GC population from the Harris (1996) catalog. We
calculate the structural parameters of each GC as follow. We assign
each cluster a mass $M_{\rm GC}$ according to
\begin{equation}
M_{\rm GC}=1.45\times10^{(4.8-M_{\rm V})/2.5},
\end{equation}
where $M_{\rm V}$ is the absolute $V$ magnitude given in the catalog,
and we used a constant mass to light ratio of 1.45 (McLaughlin 2003)
to convert the total cluster luminosity into its mass. We adopt the
observed central luminosity and core radius of each cluster listed in
the catalog and calculate the central stellar density and the core
mass. We provide each cluster with a stellar velocity dispersion
according to (McLaughlin 2000)
\begin{equation}
{\rm log}\sigma_c=0.525\times{\rm log}L-1.928
\end{equation}
where $L$ is the total cluster luminosity.

\subsection{Intermediate mass black hole populations}
\begin{figure}
\centerline{\psfig{file=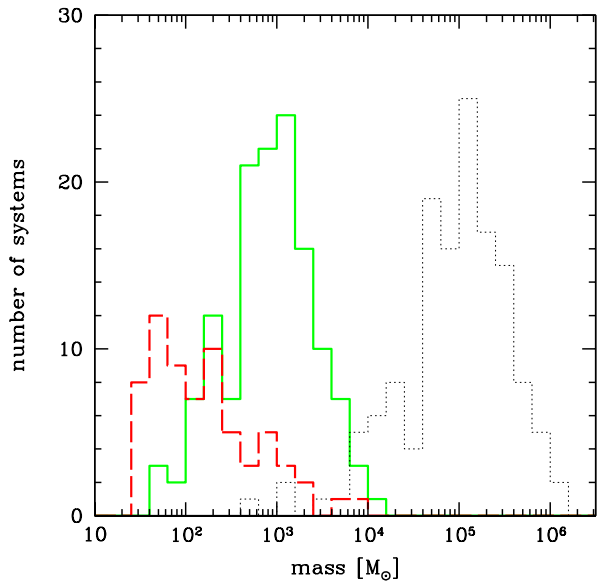,width=84.0mm}}
\caption{Dotted black histogram: distribution of GC masses (calculated using
  the Harris 1996 catalog). For each cluster we assign an IMBH with
  mass given by extrapolating the $M-\sigma$ relation (Tremaine et. al
  2002, long--dashed red histogram), or by using a recipe proposed by
  Portegies Zwart (2005) for the formation of an IMBH in a young dense
  star cluster (solid green histogram).}
\label{f1}
\end{figure}

We assign each GC a central IMBH assuming two different prescriptions
leading to a 'heavy' and a 'light' population of IMBHs. The 'heavy'
population ('H' scenario) is obtained using the recipe provided by
Portegies Zwart (2005) for the runaway collision formation of
an IMBH in a dense stellar cluster. The forming IMBH mass can be
written as:
\begin{equation}
M\simeq m_{\rm seed}+4\times10^{-3}f_c\,M_{\rm GC}\,{\rm ln}\Lambda.
\label{pzwart}
\end{equation}
Here $m_{\rm seed}=50\msun$ is the mass of the heavy star that
initiates the runaway process, ${\rm ln}\Lambda=10$ is the Coulomb
logarithm and $f_c=0.2$ is a runaway efficiency factor. As shown in
figure \ref{f1}, this prescription leads to a typical IMBH mass of the
order of $10^{-2} M_{\rm GC}$. The recipe given by equation
(\ref{pzwart}) is appropriate in the context of IMBH formation in
young dense star clusters, where the core relaxation time is shorter
than the lifetime of the most massive stars ($\sim 3$Myr). This is not
the case for GCs in the MW, which have core relaxation times
$>10^7$yrs, even though they may have suffered a significant core
expansion during their life (Binney \& Tremaine 1987). Moreover,
predicted IMBHs are quite massive, and some of them are only marginally
consistent with upper limits placed by dynamical modelling of observed
density and luminosity profiles of few selected GCs (see, e.g., van
der Marel \& Anderson 2010 for Omega Centauri). As a suitable
alternative, we explore the possibility that each cluster hosts an
IMBH consistent with a low mass extrapolation of the $M-\sigma$
relation observed in galactic bulges (Tremaine et al. 2002)
\begin{equation}
M_6=2\sigma_{70}^4,
\label{msigma}
\end{equation}
where $M_6$ is the BH mass in unit of $10^6\msun$ and $\sigma_{70}$ is
the GC velocity dispersion in units of 70 km s$^{-1}$. Figure \ref{f1}
shows that this prescription leads to a population of much lighter
IMBHs (and we will refer to this prescription as 'L' scenario). In
this case we place a lower cut to the IMBH mass: GCs that contain a
central IMBH less massive then $30\msun$ according to equation
(\ref{msigma}) are simply considered to be void of a central BH, and
do not contribute in the calculation. Note that more than 50\% of the
galactic GCs do not host an IMBH according to this recipe.


\subsection{Millisecond pulsar population}


GCs are known to host an exceptionally high number of millisecond
pulsars (MSPs). There are 22 MSPs detected in 47 Tucanae (hereinafter 47 Tuc)
and more than 30 in Terzan 5. There is, up to date, a total of $\sim$ 140 MSPs
discovered in more than 20 GCs \footnote {an updated catalog is
  available at http://www.naic.edu/~pfreire/GCpsr.html}, the 
majority of them being in tight binary systems. It is worth
to notice that ongoing surveys are only sensitive to the brightest
pulsars hosted in nearby GCs. It is therefore possible that many more
pulsars than the $\sim20-30$ currently detected are hosted by massive
GCs like 47 Tuc and Terzan 5. Dynamical modeling of dense clusters
shows that dense massive clusters in the MW can host as many as 500
MSPs (see detailed discussion in Ivanova et al. 2008a). Assuming that
$20-40\%$ of the MSPs are observable because of beaming (Kramer et al.
1998; Lorimer 2008 and references therein),
this implies $\sim100-200$ observable MSP might be hosted by such
clusters, the vast majority of them being below our current detection
threshold (Camilo \& Rasio 2005). The number of potentially observable 
MSPs in 47 Tuc estimated by Camilo et al. (2000) by extrapolating 
the observed MSPs luminosity function to the faint end is in fact $\approx 200$.
Note, however, that measurements of the integrated unresolved radio flux 
from 47 Tuc suggest that the faint undetected population might be of 
the order of just $\lsim 30$ objects (McConnell et al. 2004).  
Interestingly, more than half of the observed MSPs are in
binary systems, and are concentrated in the GC cores. Assuming a close 
correlation between low-mass X-ray binaries (LMXBs) and the MSP formation 
(van den Heuvel \& van Paradijs 1988), 
as supported by the recent observation of a MSP binary with a shut-off 
accretion disk (Archibald et al. 2009), we populate each
cluster with a number of MSPs in binaries scaling as:
\begin{equation}
N_{\rm MSP}\propto \left(\frac{\rho_0^2 r_c^3}{\sigma_c}\right)^{0.74},
\label{Nmsp}
\end{equation}
which is approximately the observed scaling between the number of LMXBs and the GC
structural parameters (Pooley et al. 2003){\footnote{Note that our assumption has a double
source of uncertainty. Firstly, the relation expressed in equation (\ref{Nmsp}) has a large 
scatter, and secondly the relation is for LMXB and not for MSPs. For example, according to equation
(\ref{Nmsp}) one would expect several MSPs in NGC 6388, where $\approx 10$ LMXB have
been detected by Chandra (Nucita et al. 2008); however, none has been found to date.}}: the central density
$\rho_0$, the core size $r_c$ and the velocity dispersion
$\sigma_c$. The term in parenthesis simply describes the formation
rate of close binaries in a dense environment: this is proportional to
the squared object density multiplied by the volume and divided by the
typical relative velocity between objects. As a study case we
normalize equation (\ref{Nmsp}) to 50 potentially observable 
(i.e., beamed toward the Earth, irrespectively of their luminosity) 
MSPs in binaries in the core of 47 Tuc. All our results can be
re-normalized by changing this number.

\section{Modelling the IMBH-binary interaction}
A large population of MSPs in binaries would be affected by the
presence of an IMBH in the GC center. Close interactions between the
binary and the IMBH would cause the tidal break-up of the binary, with
one of the two members being ejected from the cluster with high speed
and the other being captured in a very eccentric orbit (Hills 1975,
Hills 1988, Bromley et al 2006, Sesana Haardt \& Madau 2007).

\subsection{The tidal disruption cross section}
Let us start by modelling the dynamical interactions of a massive object 
embedded in a population of light field binaries. The target is an IMBH 
of mass $M$ and the bullets are binaries containing a MSP of mass $m_1=$1.4
$M_{\odot}$ and a lighter companion of mass $m_2$. 
The binary population is specified by a semimajor axis
distribution $a_b$, by the binary center of mass velocity with respect
to the IMBH $v_b$, and by the mass of the companion to the MSP
$m_2$. We denote the binary semimajor axis probability distribution by
$g(a_b)$. This distribution is derived by fitting the data presented
by Camilo \& Rasio (2005). Observations show a bimodal binary population,
characterized by half of the MSPs having a very light
($m_2\sim0.03\msun$) and very close ($a_b<0.01$ AU) companion, and the
other half being accompanied by heavier stars ($m_2\sim0.1-0.3\msun$)
in wider orbits ($a_b\sim0.02$ AU, with a long tail extending to $0.2$
AU).  We fit such distribution by (i) an asymmetric Landau profile,
peaked at 0.005 AU, in the range [0.0024 AU, 0.02 AU], plus (ii) a
Gaussian profile, centered around 0.026 AU, in the range [0.02 AU,
  0.035 AU]. We denote the center of mass velocity distribution as
$f(v)$ and we assume it follows a Maxwellian distribution with 1-D
dispersion $\sigma_c$. As we will see below, only the total mass
$m_1+m_2$ enters in the derivation of the cross section and the interaction 
rate, and since $m_2\ll m_1$, we will neglect the $m_2$ dependence 
in their computation.
The $m_2$ distribution is instead consistently sampled in the evaluation 
of the ejection velocity, which crucially depends on the mass of the MSP
companion (Section 2.3) .

The relevant lengthscale for computing the break-up cross section and 
a relative rate is then the tidal break-up radius, defined as
\begin{equation}
r_{\rm tb}=\left(\frac{M}{m_1+m_2}\right)^{1/3}a_b,
\label{rtb}
\end{equation}
We consider a close interaction to be efficient when it leads to the
break-up of the binary. In this respect, $r_{\rm tb}$ does not mark a
clear distinction between tidal break-up and binary survival, but
there is a smooth transition between the two regimes around $r_{\rm
  tb}$, described by a break-up probability function (Hills 1975)
\begin{equation}
p_{\rm tb}(r_{m}/a_b)=1-\frac{D}{175},
\label{pd}
\end{equation}
\begin{equation}
D=\frac{r_m}{a_b}\left[\frac{2M}{10^6(m_1+m_2)}\right]^{-1/3}.
\label{hillspar}
\end{equation}
Here $r_m$ is the closest approach of the binary center of mass to the
IMBH. When $r_m=r_{\rm tb}$, $p_{\rm tb}\sim0.5$, and it smoothly
approaches zero for $r_m\gg r_{\rm tb}$. To model the ejection rate of
MSPs, we need to define an interaction cross section, and an ejection
timescale. In general the interaction rate for an IMBH embedded in a
background of field objects with number density $n$ and 1-D velocity
dispersion $\sigma_c$ is given by:
\begin{equation}
\Gamma=n\,\sqrt{3}\sigma_c\Sigma,
\label{rate}
\end{equation}
and the cross section $\Sigma$ is formally defined as an integral over
the impact parameters $b$
\begin{equation}
\Sigma=2\pi\int_0^{b_{\rm max}}bdb.
\end{equation}
 Any given impact parameter is related to the closest approach $r_m$
 to the IMBH by:
\begin{equation}
b^2\simeq\frac{2GMr_m}{v_b^2},
\label{bi}
\end{equation}
where we assumed (as it is always the case for our systems) that the
cross section is dominated by gravitational focusing. The range of
$r_m$ leading to a disruption depends on $a_b$, and the focusing
depends on $v_b$. We then define an effective cross section taking into
account for the distributions $g(a_b)$, $f(v)$ and for the disruption
probability $p_{\rm tb}(r_{m}/a_b)$ by writing:
\begin{equation}
\Sigma=\frac{4\pi GM\int dv\,f(v)v_b^{-2} \int da_b\,g(a_b) \int dr_m\,r_m p_{\rm tb}(r_{m}/a_b)}{\int dv f(v) \int da_b g(a_b)}.
\label{sigma}
\end{equation}
Finally, equation (\ref{rate}) can also be rewritten as
\begin{equation}
\Gamma=n\,\sqrt{3}\sigma_c\Sigma=N\frac{\sqrt{3}\sigma_c\Sigma}{V_c}=\frac{N}{\tau},
\label{tau}
\end{equation}
defining a characteristic interaction timescale $\tau=V_c/(\sqrt{3}\sigma_c\Sigma)$.  
Here $V_c$ is the volume of the GC core, computed via
$V_c=(4/3)\pi r_c^3$, using the values of $r_c$ found in the Harris
catalog. We consider the volume of the core because most of the MSPs
are found in cluster cores, we then assume this is the volume scale
relevant for setting the dynamical timescale of the process.

\subsection{Consistent evolution of the MSP population}
We now consider the population of globular clusters given by the Harris
catalog, labelled by the index $i=1,...,146$. We explore two models
for the dynamical evolution of the binaries containing a MSP in
GCs. In both scenarios we assume that $N_0^i$ {\it potentially observable}
(i.e., with radio emission beamed toward the Earth) MSPs in binary
systems are produced in each GC over its lifetime, starting from
$t_0=-12$ Gyr up to now. In model I we consider that
all the $N_0^i$ systems were formed in a single burst at $t_0$. In
this case, the evolution of the binary population in each individual
GC is simply dictated by the tidal break--up rate, and it is described
by an exponential decay:
\begin{equation}
N^i(t)=N_0^ie^{-t/\tau^i}
\end{equation}
($\tau^i$ is the interaction timescale, as defined by equation
(\ref{tau}), which is different for each cluster).  The binary
disruption rate is
\begin{equation}
\Gamma^i(t)=\frac{N_0^i}{\tau^i}e^{-t/\tau^i},
\label{gamma1}
\end{equation}
and the number of disrupted systems in 12 Gyr is:
\begin{equation}
N_{\rm ej}^i=\int_{t_0}^0\frac{\Gamma^i(t)}{2}=N_0^i(1-e^{-t_0/\tau^i}).
\label{nej1}
\end{equation}
Note the factor $1/2$ in equation (\ref{nej1}). This is because
following the tidal break--up, one of the two component of the binary
is indifferently ejected (Sari Kobayashi \& Rossi 2010, Kobayashi et al. 2012).
It is therefore reasonable to expect that only $\sim50\%$ of the interacting MSP
will be ejected. Given a total number of binaries $N_0^i$, if the
number of ejected MSPs is $N_{\rm ej}^i$, the number of MSPs retained
in binaries is $N_{\rm ret}^i=N_0^i-2N_{\rm ej}^i$.  (Note that
MSPs that remain bound to the IMBH have characteristic coalescence
timescales that are $\lsim \, 10^8$ yr (Devecchi et
al. 2007). Formed IMBH-MSP binaries are thus expected to coalesce
before the next interaction of the IMBH with another 
binary containing an MSP.)

\begin{figure}
\centerline{\psfig{file=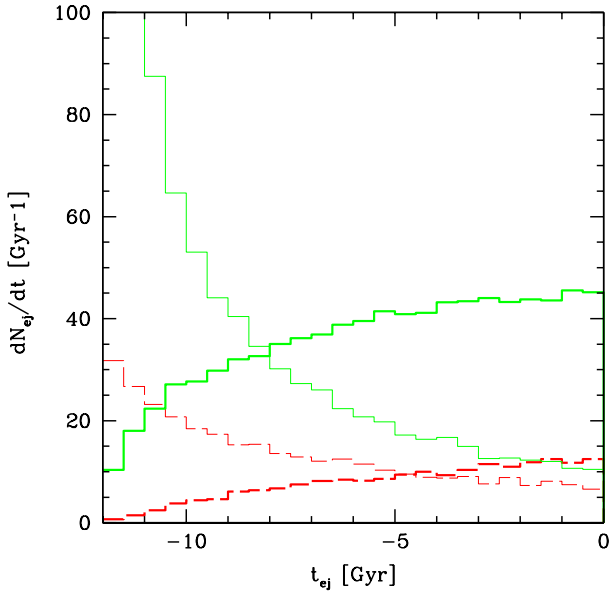,width=84.0mm}}
\caption{MSPs ejection rates for our four models. The H models (from here on) are shown with solid green lines and the L models (from here on) are shown with long--dashed red lines. Thin lines are for the MSP formation scenario I (formation burst at $t_0$) and thick lines are for the MSP formation scenario II (constant formation rate from $t_0$ to now).}
\label{f2}
\end{figure}
In model II we assume that the $N_0^i$ MSPs are formed at a constant rate from $t_0$ to now. The evolution of the population is then described by a constant formation rate term plus a loss term dependent on the break--up rate. The solution is given by:
\begin{equation}
N^i(t)=\frac{N_0^i}{t_0}\tau^i(1-e^{-t/\tau^i}).
\end{equation}
The binary disruption rate is
\begin{equation}
\Gamma^i(t)=\frac{N_0^i}{t_0}(1-e^{-t/\tau^i}),
\label{gamma2}
\end{equation}
and the number of disrupted systems in 12 Gyr is:
\begin{equation}
N_{\rm ej}^i=\int_{t_0}^0\frac{\Gamma^i(t)}{2}=N_0^i\left[1-\frac{\tau^i}{t_0}(1-e^{-t_0/\tau^i})\right].
\label{nej2}
\end{equation}
Each model provides a consistent way to calculate the total MSP
ejections from each single cluster and the rate at which the
break--ups occur. The total number of formed, ejected and retained
MSPs are simply computed by summing over the GC population:
$N_0=\sum_{i=1}^{146}N_0^i$; $N_{\rm ej}=\sum_{i=1}^{146}N_{\rm
  ej}^i$; $N_{\rm ret}=\sum_{i=1}^{146}N_{\rm ret}^i$. MSP ejection
rates for all our four models (H-I, H-II, L-I,L-II) are shown in
figure \ref{f2}, which highlights the completely different ejection
histories of the two scenarios. In models I, the rate follows a simple
exponential decay (it is actually a sum of exponential decays
because each GC has its peculiar interaction timescale $\tau^i$). In
models II, instead, the ejection rate is initially limited by the
availability of binaries, and it is a monotonic increasing function of
time.

\subsection{Ejection velocity}
\begin{figure}
\centerline{\psfig{file=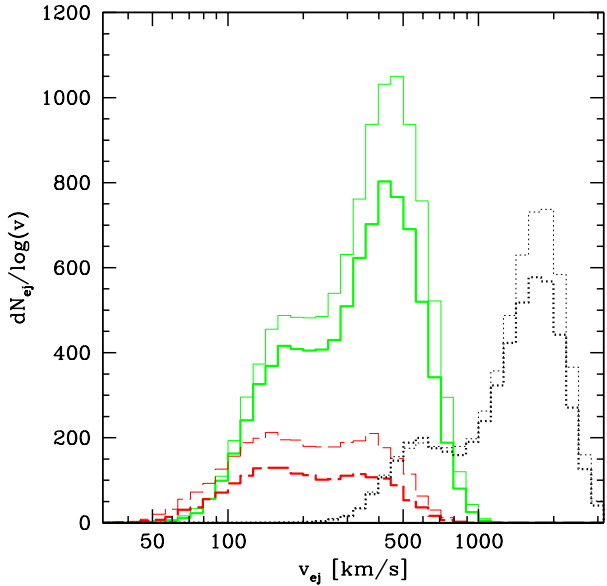,width=84.0mm}}
\caption{Velocity distribution of ejected MSPs. Linestyle as in figure \ref{f2} for the four models H-I, H-II, L-I, L-II. The dotted histograms are for a putative contaminant population of ejected MSPs from the Galactic center. Thin and thick histograms correspond to scenarios I and II for the MSP formation.}
\label{f3}
\end{figure}

After the break--up, one of the binary components is ejected with mean
velocity
\begin{equation}
v_{\rm ej}\simeq1430\,a_{b,0.01}^{-1/2}(m_1+m_2)_0^{1/3}M_3^{1/6}f_R\,\,\,{\rm km\,s}^{-1},
\label{vej}
\end{equation}
where $a_{b,0.01}$ is the binary semimajor axis in units of 0.01 AU,
$(m_1+m_2)_0$ is the binary mass in units of $1\msun$ and $M_3$ is
the IMBH mass in units of $10^3\msun$. The factor $f_R$, of order
unity, is fitted against numerical simulation and takes the form
(Bromley et al. 2006)
\begin{eqnarray}
f_R = 0.774+(0.0204+(-6.23\times10^{-4}+(7.62\times10^{-6}+\nonumber\\
 + (-4.24\times10^{-8}+8.62\times10^{-11}D)D)D)D)D,
\label{fr}
\end{eqnarray}
where $D$ is defined by equation (\ref{hillspar}). 
For any given combination of the parameters ($a_b$, $m_1+m_2$,
$M$), the distribution of ejected velocities is well approximated by a
Gaussian with mean value $v_{\rm ej}$ and dispersion $0.2v_{\rm ej}$.
The mean ejection velocity of the primary is $v_{\rm
  ej,1}=[2m_2/(m_1+m_2)]^{1/2}v_{\rm ej}$, and that of the secondary is
$v_{\rm ej,2}=[2m_1/(m_1+m_2)]^{1/2}v_{\rm ej}$. Binary MSPs are usually
found in systems with $m_2/m_1\lsim0.2$, so that their typical
ejection velocity is of the order of $300-400 {\rm
  km\,s}^{-1}$. Ejection velocity distributions are shown in figure
\ref{f3} for all our models. In all the cases, the distribution is
bimodal, reflecting the bimodality in the semimajor axis and companion
mass distributions highlighted by Camilo \& Rasio (2005). L models
produce a flat velocity distribution in the range 100-500 km s$^{-1}$, while
H models are strongly peaked around 500 km s$^{-1}$. These values are
interesting because most of the MSPs will be able to travel far in the
halo remaining bound to the Galaxy: even though the global ejection rates
are $<10^{-7}$yr$^{-1}$, we may expect an accumulation of several
hundreds of fast halo MSPs during the MW lifetime. As a possible
contaminant population, we consider a putative distribution of pulsars
ejected by the Galactic center. Since little is known about neutron
stars and MSPs in the Galactic center, we decided to model it by 
adopting the same prescriptions as for GCs. The number of MSPs in 
binaries is derived according to equation (\ref{Nmsp}) where we used 
the following parameters: $r_c=2.2$ pc, $\sigma_c=100$ km s$^{-1}$,
$\rho_0=8\times10^6/(4/3\pi r_c^3)\msun$ pc$^{-3}$. The mass of the MBH
is assumed to be $M=4\times 10^6\msun$ (Gillessen et al. 2009). As
shown in figure \ref{f3}, ejection velocities are in this case usually
higher than 800 km s$^{-1}$ (primarily because the MBH in the Galactic center
is much more massive than the putative IMBHs populating GCs), implying
that most of them will escape the MW halo, and we may expect only and
handful of contaminant fast MSPs coming from the Galactic center.
Note that according to this simple model $\gsim 1000$ binary 
MSPs are produced within 2.2pc from Sgr A$^*$ over an Hubble time. 
By comparison, X-ray counts published by Muno et al. (2003), imply 
$\sim1000$ hard X-ray sources within 20 pc from Sgr A$^*$. As the detected
space density approximately increases as the inverse square of the distance, 
we infer $\sim100$ X-ray sources within 2 pc from Sgr A$^*$.
However, only a small fraction ${\cal F}$ of them are expected to be MSP progenitor 
X-ray binaries. This implies a number of MSPs forming in binaries within 2.2pc 
around Sgr A$^*$ in an Hubble time $T_H$ of the order of $100{\cal F}T_H/T_X$; which is 
less than the fiducial number we use, as long as the average lifetime of a MSP progenitor X-ray binary 
is $T_X \gsim {\cal F}10^9$ yr. We conclude that the oversimplistic scenario adopted here 
for the Galactic center provides a reasonable estimate of the resulting
MSP contaminant population (we will further discuss contaminant populations in Section 4.3).

We should notice, at this point, that our ejected MSP population models (rates and
velocities) are derived from the MSP binary population {\it observed today}. 
However, MSP binaries might have suffered significant dynamical interactions with other stars
and binaries in the dense environment of GC cores. The typical binary-single star encounter
timescale is given by (King et al. 2003)
\begin{equation}
\tau_{\rm enc}=7\times10^{10}{\rm yr}\,\sigma_{10}n_5^{-1}R_m^{-1}M^{-1},
\label{tint}
\end{equation}
where $n_c$ is the core number density in units of $10^5$pc$^{-3}$, $\sigma$
is the velocity dispersion in units of 10 km s$^{-1}$, and $R_m$ and $M$ are the
maximum encounter approach and the total mass of the three bodies in solar
radii and solar masses respectively. For typical observed MSP binaries and dense
cluster parameters, the timescale for an encounter at $R_m\sim a$ is
$\sim 10^{10}$ yr. Hence, only perhaps few strong encounters might have 
occurred since their formation. However, strong interactions certainly
play a role in the MSP population evolution, otherwise we would not see
isolated MSPs in GCs (whereas, in fact, 40\% of them are, Lorimer 2008). 
Such interactions are neglected in the evolutionary models presented here. 
Their main result is to change the binary semimajor axis $a$ and 
the companion mass $m_2$ (in an exchange encounter), therefore affecting 
the subsequent MSP binary interaction cross section with the IMBH 
(wider binaries are easily disrupted). On the other hand, 
the ejection velocity is only mildly dependent on $a$ and $m_2$ (see equation
(\ref{vej})). As an extreme case, we tested a scenario where all the MSP 
companions have $m_2=0.5\msun$ and the $a$ distribution is three times 
wider than that described in Section 3.2. The net result is a $\gsim 50\%$ 
increase in the number of ejections with a $\sim 25\%$ decrease in the
typical ejection velocity. We therefore conclude that evolution of the 
MSP binary population will affect our results within a factor less than 
two, much smaller than the impact of changing the IMBH population 
(see red vs. green curves in figure \ref{f3}).


\subsection{Composing the puzzle}
We assign to each GC an IMBH according to one of the prescriptions
given by equations (\ref{pzwart}) and (\ref{msigma}). We then compute
$\Sigma^i$ for each GC (equation (\ref{sigma})) and the interaction
timescale $\tau^i$, assuming a population of MSPs described by a
distribution of semimajor axis $g(a_b)$ and of velocity at infinity
$f(v)$. These numbers are then plugged into equations (\ref{nej1}) and
(\ref{nej2}) to compute how many binaries are ejected according to the
two different MSP formation scenarios. To do so, $N_0^i$ is set to 50
in 47 Tuc, and it is scaled to other GCs according to equation
(\ref{Nmsp}). We then generate a Monte Carlo population of binaries
according to $g(a_b)$ and $f(v)$, we target the MBH with a
distribution of impact parameter 'at infinity' proportional to $bdb$,
and we break--up the binaries according to $p_{\rm tb}$ as given by
equation (\ref{pd}). To each disrupted binary we assign a disruption
time $t_{\rm ej}$ drawn from a probability
distribution proportional to the rates (\ref{gamma1}) and
(\ref{gamma2}), and we assign a velocity $v_{\rm ej}$ according to the
procedure described in Section 2.3 (equation (\ref{vej})).

MSPs are then ejected from GCs, or the Galactic center, and their orbits  
are integrated in the MW potential with the PSYCO\footnote{Population
SYnthesis of Compact Objects} code (Sartore et al. 2010). Orbit
integration is carried out in the following way: for MSPs ejected
from GCs, we first initialize the orbit of each GC. The initial
positions of GCs are taken from the Harris catalog, while we
assign a random velocity taken from a Gaussian distribution with
dispersion (Brown et al. 2010)
\begin{equation}
\sigma\,=\,116\,-\,0.38\,\times\,r\,\textrm{km\,s}^{-1}\,,
\end{equation}
where \emph{r} is the distance from the Galactic center in kpc.
We then follow the orbit of the \emph{i}-th GC for 12 Gyrs 
and store its position
$\textbf{r}_{\rm GC}^i$ and velocity $\textbf{v}_{\rm GC}^i$ at
each disruption time. At this point we initialize the orbit of
each MSP, its initial position and velocity being $\textbf{r}_{\rm
MSP}\,=\,\textbf{r}_{\rm GC}(t_{\rm ej})$ and $\textbf{v}_{\rm
MSP}\,=\,\textbf{v}_{\rm GC}(t_{\rm ej})\,+\,\textbf{v}_{\rm ej}$,
respectively. Both positions and velocities are defined in
a Galactocentric cylindrical frame $(R,\phi\,z)$ and the direction
of $\textbf{v}_{\rm ej}$ is random. In the case of MSPs ejected
from the Galactic center, these are initially placed at $r\,=\,2$
pc ($\sim\,r_{c}$) from the MBH, while now $\textbf{v}_{\rm
MSP}\,=\,\textbf{v}_{\rm ej}$ and is assumed to be purely radial
{\footnote{We consider purely radial velocities because the 
tidal break-up radius for a typical binary orbiting Sgr A$^*$ is of the 
order of mpc, much smaller than the initial radius of orbit integration
($r=2$ pc).}}.

\section{Results}


\subsection{Population of ejected and retained MSPs}
\begin{figure}
\centerline{\psfig{file=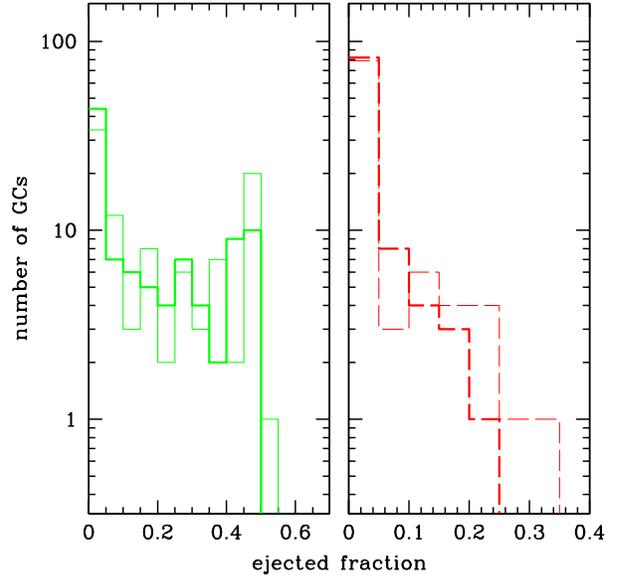,width=84.0mm}}
\caption{Fraction of ejected MSPs. Left panel: H models; right panels: L models. Note the different scale on the $x$ axis. Thin and thick histograms correspond to scenarios I and II.}
\label{f4}
\end{figure}

In figure \ref{f4} we show the distribution of the fraction of MSPs
ejected by each GC. Only clusters forming at least one MSP according
to equation (\ref{Nmsp}) are shown. In the L scenario, the vast
majority of the clusters retain all the MSPs because there is no IMBH
in their center, or the IMBH mass is so small that its tidal break--up
cross section is negligible. Only $\sim20$ GCs eject a significant
fraction of their MSPs. In the H scenario still $\sim30\%$ of the
clusters retain all their MSPs, but several of them eject a
significant fraction of systems. Note that here we plot the fraction
of {\it ejected} MSP, and only $50\%$ of the disrupted binaries result
in the MSP ejection. A fraction of ejected pulsars of 0.5 then means
that the IMBH has disrupted all the binaries hosting MSPs, and the other
0.5 fraction of MSPs remained bound to the IMBH. This scenario is
consistent with results of 3-body simulations (Devecchi et
al. 2007).  Further evolution of newly formed IMBH--MSP binaries is
dictated by i) hardening due to three body interactions with
other cluster stars and ii) gravitational wave emission. These two 
processes generally lead to a rapid coalescence ($\sim\,10^8$ yr) of the binaries.

\begin{figure}
\centerline{\psfig{file=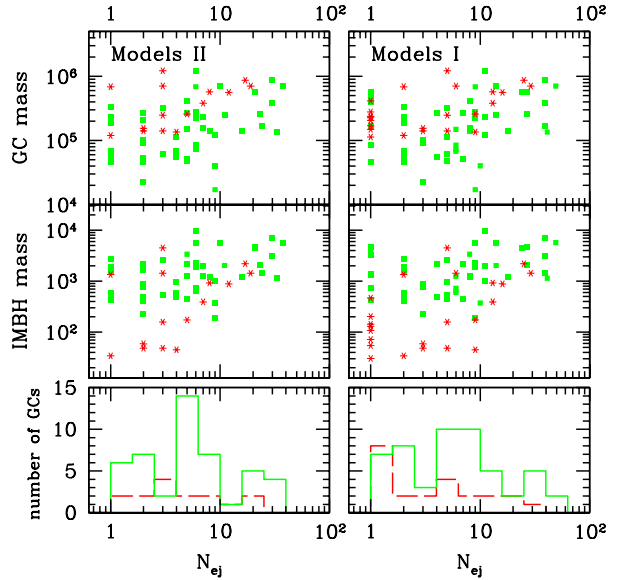,width=84.0mm}}
\caption{Statistics of ejected MSPs. The different panels represent the number of ejected MSPs versus the mass of the GC (top) and versus the mass of the IMBH (middle), and the distribution of the number of ejected MSPs from each individual cluster (bottom). Left panels are for scenario II and right panels are for scenario I. In each panel, red asterisks are for the L IMBH population model and filled green squares are for the H IMBH population model. In the lower panels, solid green and long--dashed red histograms are for H and L models respectively.}
\label{f5}
\end{figure}

The contribution of each GC to the ejected population of MSPs is shown in 
figure \ref{f5}. If the IMBH mass is assigned according to the $M-\sigma$ 
relation, then only $\sim20$ GCs contribute to the halo MSP population 
ejecting at least one object.  Interestingly, BHs with masses as low as 
100$\msun$ can eject $\sim10$ MSPs during an Hubble time. In the H scenario, 
$\sim50$ GCs eject at least 1 MSP. The total numbers of ejected and retained 
MSPs are given in table \ref{tab1}. The number of fast MSPs ejected in the 
halo spans almost an order of magnitude, from $\sim100$ (L-II model) to 
$\sim600$ (H-I model), this number can be as high as $\sim1000$ if the ejection
probability for the MSP is higher than 0.5{\footnote {Some author in fact 
(i.e., Bromley et al. 2006) suggest that in the tidal break--up process, 
if $m_2\ll m_1$, then the probability of ejecting the more massive object is 
substantially higher than $0.5$. This is always the case for our systems, 
since the MSP companion is always a very low mass main sequence star. However,
here we preferred to be conservative and assume equal probability for the 
ejection of the MSP and the low mass companion.}}. The number of retained 
binaries containing MSPs is instead in the range $800-1700$ for the different scenarios. 
It is therefore very unlikely that a population of IMBHs would completely 
devoid MSPs in binaries residing in GCs. In the most extreme case, 
$\sim50\%$ of the formed binaries are disrupted by scattering with the IMBH. 
This is because, in general, the characteristic size of the GC core is 
much larger than the typical disruption cross section $\Sigma$, and 
even assuming that the 'loss cone' is always full (i.e. the nuclear 
relaxation timescale is much less than the Hubble time, which is always 
the case for the GC cores), the IMBH does not have enough time to 
tear apart all the binaries.

\begin{figure}
\centerline{\psfig{file=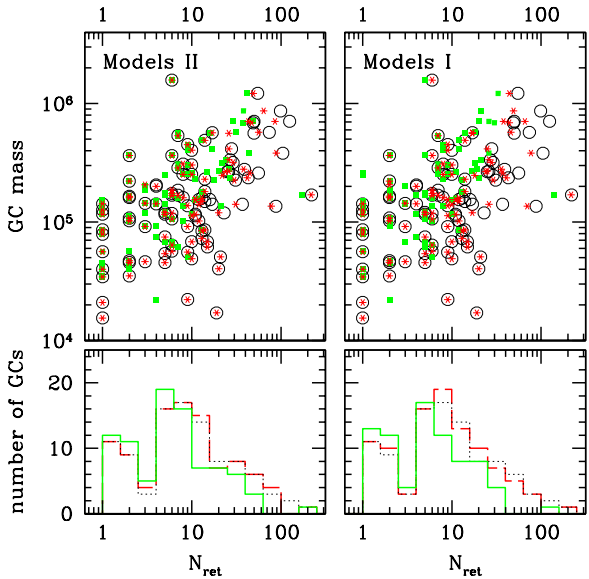,width=84.0mm}}
\caption{Statistics of retained MSPs. In the top panels we plot the number of MSPs in binaries left in the GCs today versus the mass of the GC, while in the bottom panels we plot the distribution of the number of retained MSPs in each individual cluster. Left panels are for scenario II and right panels are for scenario I. In the top panels, red asterisks are for the L IMBH population model and filled green squares are for the H IMBH population model, while black empty circles mark the total number of generated MSPs in binaries ($N_0^i$) in each cluster. In the lower panels solid green and long--dashed red histograms are for H and L models respectively; the thin dotted histogram represent the population of generated MSPs.}
\label{f6}
\end{figure}

As a 'consistency check', we plot in figure \ref{f6} the number of 
surviving MSPs in binaries for each GC. Most of the clusters retain 
the majority of their MSPs. In particular, even in our H scenario, 
we find that 47 Tuc would still host $\sim 25$ observable MSPs in binaries
in its core, while $\sim 50$ retained MSPs are expected in Terzan 5; 
both numbers are consistent with observations.

\begin{table}
\begin{center}
\begin{tabular}{c|ccccc}
\hline
${\rm MODEL}$  & $N_0$ & $N_{\rm ret}$ & $N_{\rm ej}$ & $N_{\rm ej,obs}({\cal A})$ & $N_{\rm ej,obs}({\cal B})$\\
\hline
   H-I&    1929&      789&   570&  64& 18\\
   H-II&   1929&     1047&   441&  50& 11\\
   L-I&    1929&     1597&   166&  25&  6\\
   L-II&   1929&     1706&   103&  14&  3\\
\hline
\end{tabular}
\end{center}
\caption{MSPs in the four different models (first column). We show the total numbers of produced MSPs in binary systems ($N_0$, column 2), the number of retained MSPs following binary disruption ($N_{\rm ret}$, column 3), and the number of ejected ones ($N_{\rm ej}$, column 4). Columns 5 and 6 are the number of MSPs ejected by IMBHs residing in GCs into the MW halo that will be detectable with SKA in configuration ${\cal A}$ and ${\cal B}$, according to simulations described in Section 5.}
\label{tab1}
\end{table}

\subsection{Properties of the ejected MSPs}

As stated in the previous section, since MSPs usually have a 
low mass companion, their typical ejection velocities are not extremely 
high, and only a small fraction of them has $v>500$ km s$^{-1}$; 
thus, they accumulate in the halo basically for an Hubble time, resulting 
in a consistent population of tens-to-hundreds fast halo MSPs. 
Therefore, in both H and L scenarios most MSPs remain gravitationally bound to the Galaxy, 
building up a halo population that extends up to $\sim\,100$ kpc from the 
Galactic center. Typical velocities of these objects are $\sim\,200\,\rm km\,s^{-1}$, 
which imply proper motions of $\sim\,0.1\,-\,10\,\textrm{mas\,yr}^{-1}$, 
i.e. well within the capabilities of present and upcoming instruments 
(especially the SKA). A small fraction of MSPs is in unbound orbits and 
thus can reach distances of several Mpc.

On the other hand, MSPs ejected from the Galactic center are almost all 
gravitationally unbound, given the higher ejection velocity induced by 
Sgr A$^*$ compared to the IMBHs. As a consequence, they do not accumulate 
in the MW halo, travelling at 
very large radii, up to $\sim\,10$ Mpc.  This is particularly important, 
because it means that such possible 'contaminant' population would be 
essentially undetectable (these MSPs are too far and thus too dim) and 
can be neglected. For the same reason, hereafter we 
shall consider the contribution of MSP within $d_{max}\sim100$ kpc from the Sun.

\subsection{Possible sources of contamination}
The detection of a fast halo MSP population may provide indirect evidence 
of IMBHs in GCs. However all possible source of contamination have to be
taken into account. 

Firstly, the Galactic disk and bulge
are known to house $\sim 30000$ MSPs (Lorimer 2005, Story et al. 2007),
born via standard binary evolution channels (Bhattacharya \& van den Heuvel 1991).
Although by far larger than our predicted fast halo MSP population,
there are several considerations that mitigate the contaminant effect
of disk MSPs. MSPs are 'recycled' objects (van den Heuvel 1995), and are 
not affected by strong natal kicks as the ordinary ones. Owing to their unperturbed binary 
evolution, these MSPs have lower spatial velocities than halo MSPs, with typical 
transverse velocities of $\sim 80 - 200\, \rm km\,s^{-1}$ (e.g. Gonzalez et al. 2011). 
As a consequence, their spatial distribution is flattened towards 
the Galactic disk, with scale-height of $\sim 0.5$ kpc (Story et al. 2007). 
Moreover, the vast majority of disk MSPs have a binary companion 
($\approx 75\%$, Lorimer 2008), whereas the break-up mechanism implies 
that halo MSPs are all isolated.
In any case, given the large disk MSP number, we need to take them into account in
simulating observations, in order to quantify the effective separability
of the two populations. 

Another source of contamination might be due to MSPs released in the halo by
cluster evaporation and disruption. It is in fact likely that observed GCs 
were more massive in the past, undergoing a slow process of dynamical evaporation
and tidal stripping (e.g. McLaughlin \& Fall 2008). However,  MSPs and their massive 
progenitors are among the heaviest GC objects, and preferentially segregate 
in the core, rather than being ejected. Following the same line of arguments,
a substantial number of GCs
might have completely disappeared by now, releasing their stellar content,
including putative MSPs, in the halo. However, this release mechanism does not require 
any strong dynamical interaction, and the MSPs would naturally retain their donors. 
We also notice that clusters prone to evaporation and disruption
are the less dense and massive ones, minimizing (i) the efficiency of MSP production 
in first place; and (ii) the probability of dynamical interactions depriving the MSPs 
of their donors. We therefore expect the contamination from halo MSPs produced in completely evaporated clusters to be minor, and mostly composed of MSPs in binary systems.

A further minor contaminant population might come from
MSPs directly formed in the halo. The MW halo, in the range 1--40 kpc from the 
Galactic center, consists of $\sim 3\times10^8\msun$ of old stars (Bell et al. 2008). 
Even assuming the same formation rate as in the disk, a population
of MSPs directly formed in the halo would count $\sim 100$ objects. This would
be a significant contaminant in our L scenario (see Table \ref{tab1}). 


In addition to contaminant populations, we also have to be careful in considering 
other possible formation channels of fast halo MSPs not requiring IMBHs. 
The formation of these halo objects must necessarily be dynamical, involving some close 
interaction with other compact objects. Such dynamical interactions 
are far too rare in ordinary stellar environment (galactic disk and bulge), 
and any other formation process must take place in a dense environment, i.e.  GCs or 
the Galactic center. We already discussed the possibility of ejecting MSPs 
from the Galactic center, and how the resulting population would be 
significantly different with respect to the one ejected by IMBHs 
(and, in any case, hardly detectable).  

We examine here possibility of ejection following the
binary break-up due to binary-single or binary-binary interactions.
According to equation (\ref{tint}) MSP binaries might have suffered 
few strong encounters with single objects in their lifetime. 
For typical parameters, MSP binaries are extremely hard
(i.e., specific binding energy to average GC star specific kinetic energy, 
$\epsilon_b/\epsilon_k$, $ \gg 1$), and the most probable outcomes of a 
three body interaction are either a flyby with the binary getting harder (unless
the intruder is a stellar BH, see below), or an exchange of companion star:
in any case, MSP ejection is very unlikely.
Binary-binary interactions are far more complex, and therefore difficult 
to address. In such encounters the MSP binary is usually the 'harder' in the pair.
Therefore, also in this case, it tends to get harder while the other binary 
gets softer or ionized. The typical interaction outcome is 
a simple flyby or a breakup of the softer pair (Bacon, Sigurdsson \& Davies 1996).
Companion exchanges are also possible, whereas individual MSP ejection is unlikely.
Close encounters also perturb the MSP binary orbital elements (semimajor axis and
eccentricity), possibly leading to the 
coalescence of the pair. In fact, maybe half of the observed individual binaries 
in Terzan 5 and 47 Tuc might be the endproduct of dynamically induced 
mergers (Ivanova et al. 2008a). In this case, no appreciable kick is expected, 
and the MSP is retained in the cluster. In general, it is likely that individual 
MSPs will be located approximately where they formed (in cluster cores), 
with some of them displaced to the cluster halo (Ivanova et al. 2008b). In fact, 
most of the single MSPs found in clusters are observed in the core, 
which would be unlikely if the typical outcome of close encounters is 
ejection at high velocity.

Lastly, MSP binary tidal break-up may occur due to interactions with 
stellar BHs.  The ejection velocity is in fact weakly dependent on 
the BH mass (see equation \ref{vej}), meaning that the survival of few stellar BHs in
GC centers, may mimic the effect given by IMBHs. However, stellar BHs
are thought to segregate efficiently at the center of GCs, forming
binaries that interact with each other, resulting in a quick ejection
of basically all the stellar BHs from the clusters. Simulations
carried by O'Leary et al. (2006) show that basically all the stellar
BHs would be ejected in less than a Gyr. We run a test case where we
assume a single 10$\msun$ BH surviving in each cluster. In this case
only 10-20 MSPs are ejected, and typical ejection velocities are much
lower, in the range $50-200$ km s$^{-1}$. Nevertheless, in the unlikely
eventuality that all the GCs retain several ($>10$) stellar BHs, they
would produce a population of ejected MSPs at least comparable to the
L model. However, being ejection velocity usually smaller, the
resulting MSP distribution would be more concentrated toward the
Galactic center, tracing closely the GC distribution in the MW
potential.  

Therefore, from an observational prospective, the most important contaminant 
population is that of MSPs born in the Galactic disk through the standard 
binary evolution channel.
Since disk MSPs are much more numerous than halo MSPs, they are likely to dominate 
the distribution of observed objects, and should be taken into account
before claiming the detectability of the fast halo population. In the next chapter we 
will present a method to characterize halo MSPs and differentiate them from Galactic disk MSPs.

\section{The observability of halo MSPs}

The detection of a population of fast isolated halo MSPs would be a strong, 
albeit indirect, evidence of the existence of IMBHs in GCs. In the previous section we described 
the properties of such population; we discuss here prospects for its future detection 
by means of upcoming radio facilities such as the SKA.

\subsection{Simulations of the Galactic MSP population}
According to Smits et al. (2009), the SKA will potentially detect up to $6000$ MSPs 
from the Galactic disk. This number exceeds that of MSPs which, according to our model, 
are likely to populate the halo, even in the most favorable case.

Thus, we simulate the radio properties of halo MSPs with the {\tt PSRPOP} code
(Lorimer et al. 2006) and compare them with those of Galactic disk MSPs, in order 
to find an efficient identification method. We assume that both populations have the 
same intrinsic properties: pseudoluminosities follow a power-law distribution with 
slope -1 between 0.1 and 100 mJy at 400 MHz (Lyne et al. 1998). 
Pulse periods are drawn from a lognormal distribution with mean 1 and
dispersion 0.2 in units of $\log P$, where $P$ is in milliseconds.
Pulse widths are assumed to be $20\%$ of the pulse period.
Finally the spatial distribution of disk MSPs is modelled according to 
Faucher-Gigu\'ere and Loeb (2009), with scale-height 0.5 kpc and scale-length 
4 kpc, whereas that of halo MSPs in determined by direct orbit integration
in the Galactic potential as detailed in Section 3.4. 
In all runs, we simulate 30000 disk MSPs and a number of halo MSPs  
taken from Table \ref{tab1}. 

\begin{figure*}
\begin{tabular}{cc}
\includegraphics[scale=0.4,clip=true,angle=0]{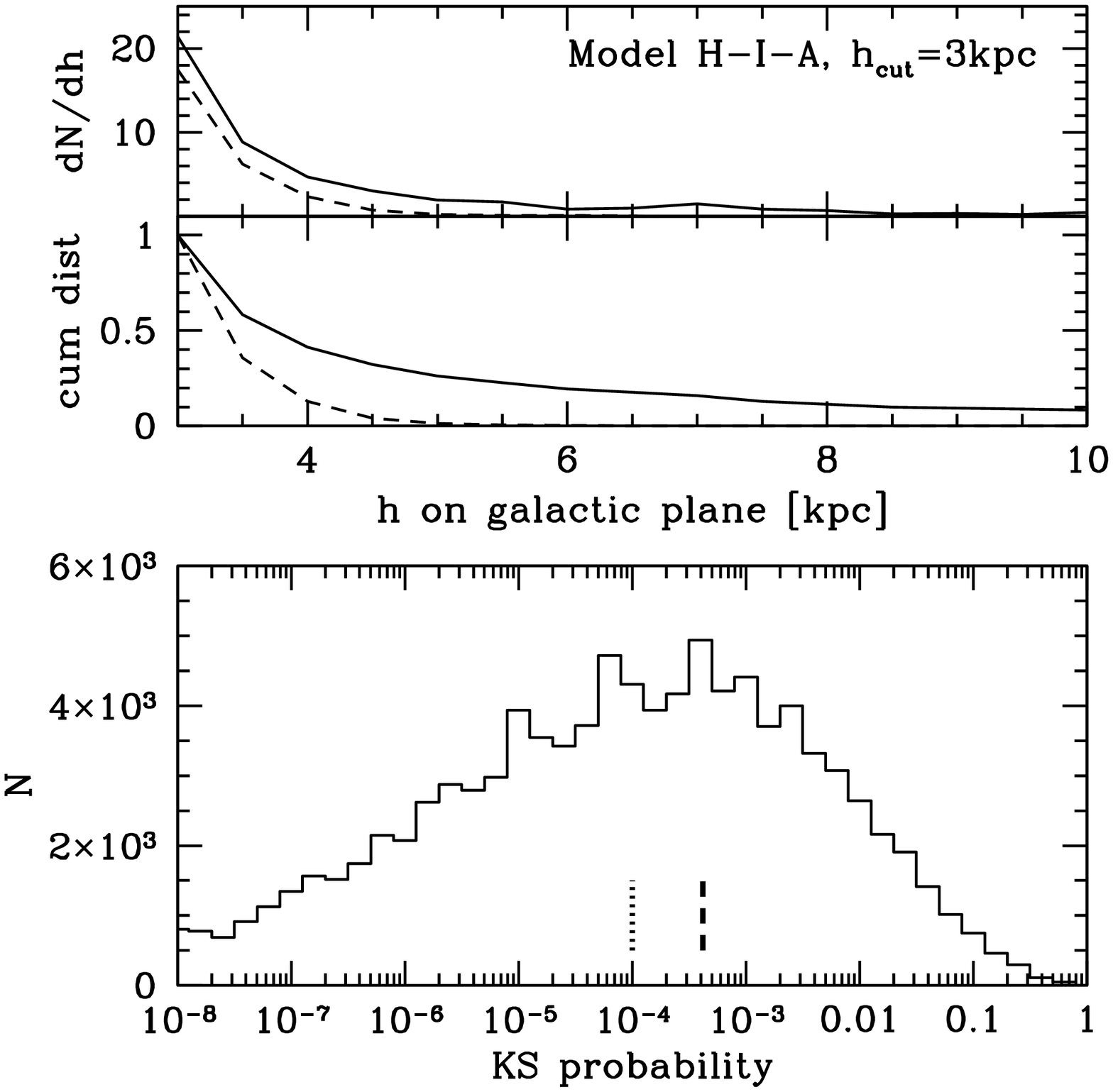}&
\includegraphics[scale=0.4,clip=true,angle=0]{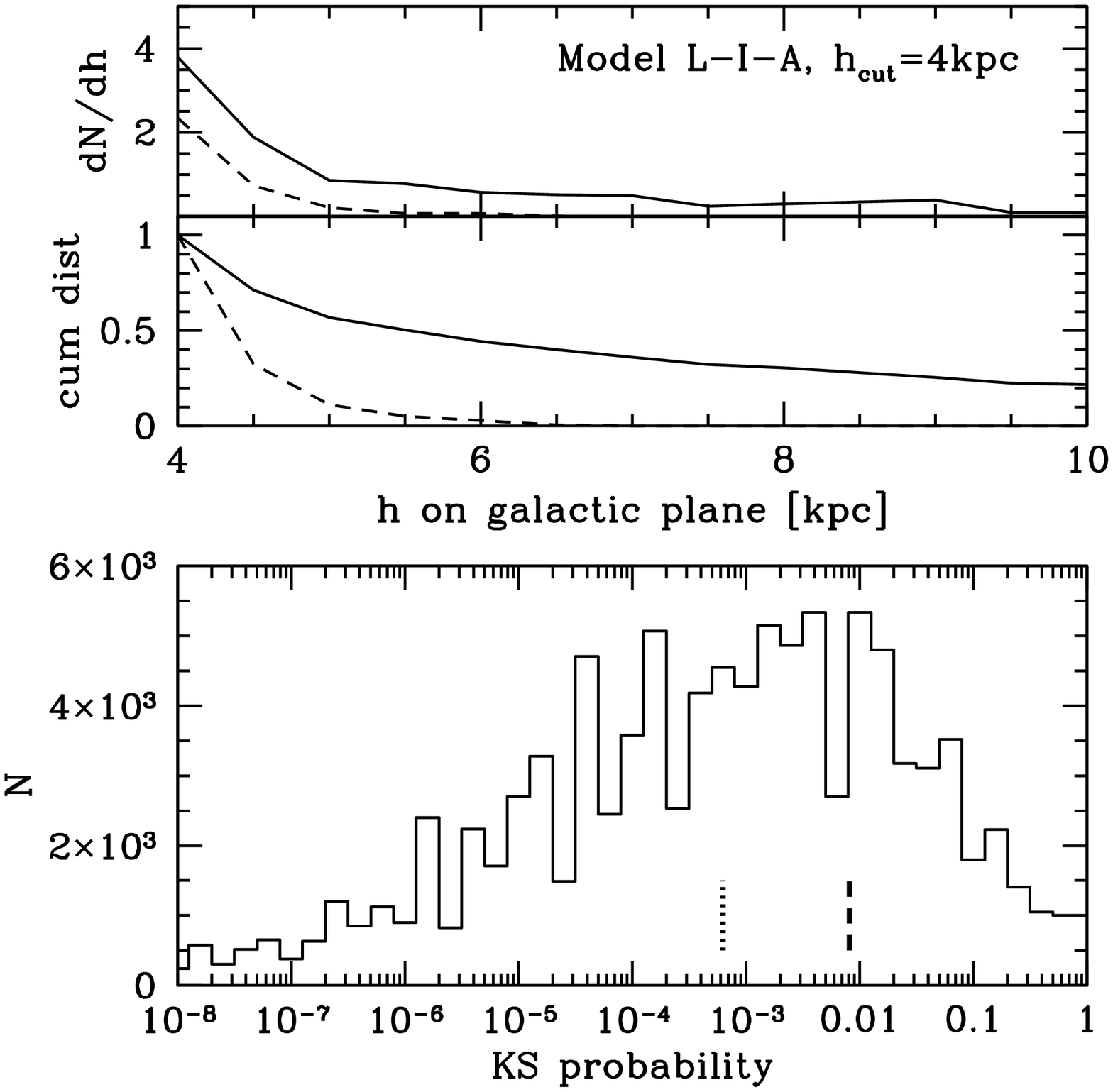}\\
\end{tabular}
\caption{Distribution of observable MSPs and $P_{\rm KS}$ for selected models. In the upper panels
we plot the tail of the observed distribution at $h>h_{\rm cut}$ (top) and the cumulative distribution
starting from $h=\infty$ down to $h_{\rm cut}$. Dashed lines are for the {\it disk} population and 
solid lines for the {\it disk+halo} population. The lower panels show the distribution of $P_{\rm KS}$
that a {\it disk+halo} Montecarlo sample is erroneously identified as being drawn by the {\it disk}
population only for $10^5$ random trials. The vertical ticks represent the mean (dashed) and the
median (dotted) of the distributions.}
\label{f8}
\end{figure*}

We model SKA observations following Smits et al. (2009). Assuming a minimum 
signal-to-noise (S/N) ratio, then the limit flux density $S_{\rm lim}$ to which
the instrument is sensitive can be obtained with the so-called radiometer equation

\begin{equation}
S_{\rm lim} = \frac{S/N\, (T_{\rm sys} + T_{\rm sky})}{G (n_p\,t_{\rm obs}\,B)^{1/2}} \Big(\frac{W}{P - W}\Big)^{1/2} {\rm mJy},
\label{eq_radiom}
\end{equation}

\noindent where $T_{\rm sys}$ and $T_{\rm sky}$ are respectively the system noise 
temperature and the sky temperature in Kelvin, $G$ is the telescope gain in 
$\rm K\,Jy^{-1}$, $n_p$ is the number of polarizations, $t_{\rm obs}$ is the 
integration time in seconds, $B$ is the bandwidth in MHz and $P$ and $W$ 
are the pulse period and width of the pulsars in seconds respectively.
We adopt the following values for the parameters: 
$T_{\rm sys} = 30$ K, $n_p = 2$, $t_{\rm obs} = 1800$ s, and $B = 512$ MHz.
For the gain we assume two possible values, relative to two different  
configurations of the SKA, one with 3000 15m dishes and gain 
$G \sim 140\, \rm K\,Jy^{-1}$ (case ${\cal A}$) and another configuration 
accounting only the inner core of the array with gain a factor 5 lower 
(case ${\cal B}$).

The average number of detectable disk MSPs is $\sim 4300${\footnote{The 
difference with the figures quoted by Smits et al. 2009 (i.e. $\sim$ 6000 objects)
comes from the different assumptions adopted for the (loosely constrained) disk population. We 
used the period and luminosity function by Lyne et al. 1998, combined with the spatial distribution
by Faucher-Gigu\'ere and Loeb (2009), whereas Smits at al. consider a slightly different 
population model by Cordes and Chernoff (1997).}} and $\sim 1150$ for 
SKA configuration ${\cal A}$ and ${\cal B}$ respectively. For comparison, 
detectable halo MSP are always less than 100; numbers for the different 
models are given in Table \ref{tab1}. For each simulation we store the
position and velocity vectors of all detectable MSP, which we use to 
analyze their observational properties. To smooth the distributions
we average over 1000 random realizations of the disk population and 
20 full numerical integrations of each halo population (models H-I, H-II, L-I,
L-II) in the Galactic potential.  

\subsection{Statistical analysis}

\begin{table*}
\begin{center}
\begin{tabular}{c|cccccccc}
\cline{2-9}
&  \multicolumn{4}{|c|}{$h_{\rm cut}=3$kpc} &  \multicolumn{4}{|c|}{$h_{\rm cut}=4$kpc}\\
\hline
${\rm MODEL}$ & $N_{\rm disk}$ &  $N_{\rm disk+halo}$ & mean $P_{\rm KS}$ & median $P_{\rm KS}$ & $N_{\rm disk}$ &  $N_{\rm disk+halo}$ & mean $P_{\rm KS}$ & median $P_{\rm KS}$\\
\hline
   H-I-${\cal A}$ & 27.11 & 51.45 & $4.31\times10^{-4}$ & $10^{-4}$ &  3.47 & 21.22 & $9.53\times10^{-5}$ & $4.90\times10^{-6}$\\
   H-II-${\cal A}$ & 27.11 & 48.71 & $2.17\times10^{-3}$ & $6.31^{-4}$ &  3.47 & 18.87 & $6.9\times10^{-4}$ & $2.51\times10^{-5}$\\
   L-I-${\cal A}$ & 27.11 & 39.65 & $3.22\times10^{-2}$ & $2.51^{-2}$ &  3.47 & 13.17 & $7.99\times10^{-3}$ & $6.31\times10^{-4}$\\
   L-II-${\cal A}$ & 27.11 & 33.76 & $0.26$ & $0.25$ &  3.47 & 8.16 & $5.52\times10^{-2}$ & $1.58\times10^{-2}$\\
   H-I-${\cal B}$ & 5.93 & 12.78 & $7.47\times10^{-2}$ & $3.98\times10^{-2}$ & 0.76 & 5.76 & $2.66\times10^{-2}$ & $1.58\times10^{-3}$\\
   H-II-${\cal B}$ & 5.93 & 12.73 & $4.55\times10^{-2}$ & $1.58\times10^{-2}$ & 0.76 & 6.16 & $9.50\times10^{-3}$ & $2.55\times10^{-4}$\\
    L-I-${\cal B}$ & 5.93 & 10.83 & $6.59\times10^{-2}$ & $1.58\times10^{-2}$ & 0.76 & 5.21 & $8.04\times10^{-3}$ & $1.58\times10^{-4}$\\
    L-II-${\cal B}$ & 5.93 & 10.03 & $0.17$ & $0.16$ & 0.76 & 3.81 & $3.43\times10^{-2}$ & $3.98\times10^{-3}$\\
\hline
\end{tabular}
\end{center}
\caption{KS probability for all the eight (four halo MSP generation models times two SKA configurations) 
investigated scenarios. The first column gives the model label; columns two-to-five give the average number of
observable {\it disk} MSPs, the average number of observable {\it disk+halo} MSPs and the mean and the median
KS probability for $10^5$ Montecarlo draws (see text for description) assuming $h_{\rm cut}=3$ kpc. 
Columns five-to-nine give the same quantities for the $h_{\rm cut}=4$ kpc case.}
\label{tab2}
\end{table*}

With the results of our simulations in hand, we seek for a suitable discriminant 
between halo and disk objects. Even though the two populations are quite distinct, this is 
not a trivial task. One obvious discriminant is the binary nature: halo MSPs are isolated, whereas
the vast majority of observed disk MSPs (about $75\%$) are in binaries. However, there are still
far more isolated disk MSPs than halo MSPs, and further 
discriminating properties are necessary. Unlike the case of Hypervelocity
stars in the MW halo (Bromley et al. 2006, Sesana et al. 2007), radial velocities 
cannot be directly measured, and in any case they would be mostly in the range $50-500$ km s$^{-1}$, 
providing just  a little (if any) discriminant power. Transverse velocities can be measured 
from the parallax; however, expected values of $0.1-10$ mas/yr are the same for both
populations. Instead of velocity, it turns out that a better discriminant between the two population 
is the spatial location of the objects, in particular the distribution of the projected 
distance $h$ of the MSPs from the disk plane. The good thing about $h$ is that it can 
be easily measured by SKA. In fact, since MSPs sky locations are known almost exactly,
the accuracy in the $h$ measurement is directly related to the measurement of the 
MSP distance, which for SKA is expected to be better than $\sim 10\%$ at 10 kpc 
(Smits et al. 2011) thanks to very precise parallax determination.
Halo MSPs should reflect the distribution of 
the parent population of GCs and thus many of them are likely to be found 
at large distances from the disk. The spatial distribution of disk MSP 
is instead expected to be extremely flattened in the galactic plane, decaying exponentially
at large distances. To quantify this concept, for each GC ejected MSP 
population and SKA configuration we extract two distinct distributions:
\begin{itemize}
\item the {\it disk} only distribution of
detectable MSPs with disk plane distance $h>h_{\rm cut}$, with $h_{\rm cut}=3, 4, 5$ kpc;
\item the {\it disk$+$halo} distribution of
detectable MSPs, for the same $h_{\rm cut}$ values.
\end{itemize}
Examples of such distributions are given in the upper panels of figure
\ref{f8} and numbers are quantified in table \ref{tab2}. The figure highlights that 
the chance of getting an MSP at $h>5$ kpc belonging to the {\it disk} population 
is almost nil (dashed lines); conversely, 
when the {\it disk+halo} population is considered, the distributions show a 
long tail extending to $h>10$ (solid lines). This is particularly clear by looking 
at the normalized cumulative  distributions; such distributions are built starting nominally 
at $h=\infty$, down to a specific value $h_{\rm cut}$, which, as stated above, we place 
at 3, 4 and 5 kpc from the galactic plane. The two reported examples highlight the extreme 
different behavior of the cumulative distributions, which is the starting point 
of our statistical tests.

The number of MSPs found at large $h$ is already a good indicator of the presence of 
halo MSPs, as shown in table \ref{tab2}. Here, columns 6 and 7 report the average number
of MSPs found at $h>4$ kpc for the {\it disk} and the {\it disk+halo} distribution respectively.
The paucity of {\it disk} pulsars indicates that the detection of a sizable number of
MSPs at large $h$ would be a strong hint of a substantial presence of halo MSPs ejected
by GCs. To make our analysis more quantitative,
we perform, for each ejection model (H-I, H-II, L-I, L-II) SKA configuration
(${\cal A}, {\cal B}$) and $h_{\rm cut}$ (3, 4, 5 kpc), a simple Kolmogorov-Smirnov (KS) 
test on the $h$ distribution predicted by the two populations ({\it disk} and {\it disk+halo}). 
We proceed as follow. We assume that SKA observes a {\it disk$+$halo} population of MSPs. 
To simulate this, we draw a number of observed MSPs, $N_{\rm obs}$, from a Gaussian
distribution with mean value equal to $N_{\rm disk+halo}$ and variance equal to 
$\sqrt{N_{\rm disk+halo}}$; values of $N_{\rm disk+halo}$ are given in table \ref{tab2}
for the different cases. We then contrast the observed 
$h$ distribution to that predicted by the {\it disk} population only and compute the 
KS indicator $P_{\rm KS}$. This namely gives the probability that our {\it disk$+$halo} drawn sample 
is erroneously recognized as being drawn from the {\it disk} only distribution. We repeat 
the procedure 100000 times. The result depends on the specific sample, and examples of the 
$P_{\rm KS}$ distribution for two selected models are shown in the
bottom panels of figure \ref{f8}. Even though the $P_{\rm KS}$ distributions have a large dispersion, 
their mean and median values are usually small, indicating that a characteristic realization of 
the {\it disk$+$halo} distribution can be identified with high confidence. Results of course depend
on the underlying halo population and SKA configuration, and also on the adopted value of 
$h_{\rm cut}$. If $h_{\rm cut}$ is too small, then the overall distribution is dominated 
by the disk population, and the large $h$ tail contribution given by halo MSPs
becomes marginal, making the KS test ineffective. On the other hand, if $h_{\rm cut}$ is too large,
then the number of detectable MSPs is very small, making the test ill defined (at least 4-5 
datapoints are needed for a trustworthy KS test). We find an $h_{\rm cut}$ range of 3-5kpc to be 
appropriate. Results are given in table \ref{tab2}. Assuming $h_{\rm cut}=3$ kpc, median $P_{\rm KS}$ are
in the range $10^{-4}-3\times10^{-2}$ for all models but L-I-${\cal A}$ and L-I-${\cal B}$, meaning
that the halo population is detected at least at a 97\% confidence level. When $h_{\rm cut}=4$ kpc,
numbers of detected MSPs get smaller, but the cumulative distributions are even more distinct 
(see, e.g., the upper right panels of figure \ref{f8}), resulting in $P_{\rm KS}<0.01$ for all
models, i.e., in 
a detection confidence greater than 99\%. $P_{\rm KS}$ values get again bigger for $h_{\rm cut}=5$ kpc
(not shown in the table), because the extremely low number of detected MSPs undermines the effectiveness
of the KS test. Notice that for some specific model we obtain $P_{\rm KS}<10^{-4}$, resulting in 
a detection confidence $>99.99$\%; moreover, results are extremely encouraging even assuming
the reduced SKA ${\cal B}$ configuration. In all our tests we assumed an underlying disk population of
30000 MSPs. This number is somewhat uncertain, but the power of the KS test lies in the fact
that it compares cumulative normalized distributions, which remain quite distinct even adding
a substantial population of disk MSPs. Tests on models with 60000 disk MSPs 
give somewhat larger (factor of $\approx2$) $P_{\rm KS}$, leaving our main results unchanged.
We stress that in this analysis we considered all the disk MSPs, including those in binaries
(the vast majority). If we restrict the disk sample to isolated MSPs, then the number of observable 
systems is expected to drop by a factor of four, making the halo population even more dominant at high
$h_{\rm cut}$. This, in turn, will strengthen our identification criterion.

\section{Conclusions}
Inspired by the detection of several MSPs in binaries in the core of 
Galactic GCs, we investigated their dynamical interaction with putative 
IMBHs lurking in the center of the same GCs. As a consequence
of three body interactions, MSPs are ejected in the Galactic halo with
velocities (with respect to the host GC) up to several hundreds km s$^{-1}$,
sufficient to distribute them into the outer halo,
but not to escape the Galaxy potential, allowing the formation of 
a substantial population of fast halo MSPs. The detection of such 
population would be a strong element, albeit indirect,
in favor of the presence of IMBHs in the center of GCs.

We investigated four ejection scenarios, involving two different IMBH 
populations and MSPs formation histories. We considered a high -H- and a low -L- 
IMBH mass functions, consistent with the runaway model of Portegies Zwart et al. (2005) 
and with the low mass end extrapolation of the $M-\sigma$ relation respectively. 
We also assumed two different MSPs formation scenarios, one in which 
MSPs are simultaneously generated in a single burst 12Gyr ago, and one with 
a constant MSP formation rate along the whole Galaxy history. In both cases,
the MSP population in GCs was normalized to match the observed numbers 
of MSPs in close binaries in the clusters Terzan 5 and 47 Tuc, and then scaled 
to all other GCs according to their structural parameters. The investigated scenarios 
predict between 100 and 600 fast MSPs wandering in the MW halo, a population
that can in principle be detected with future radio surveys. We emphasize here that 
all the considered models are consistent with current MSP observations in GCs; that is, 
even considering efficient binary break-up and MSP ejection, GCs retain
a substantial population of binaries containing MSPs, as observed today.

In the spirit of using fast halo MSPs as a probe of IMBHs in GCs, 
we checked that such population cannot result from alternative 
channels. Halo MSPs can be in principle produced by other dynamical ejection 
mechanisms, such as binary-binary or three body interaction with stellar BHs,
or can be relics of the natural evolution of the old halo stellar 
population. Both channels looks inefficient in producing a sizable
population of halo MSPs, unless several tens of stellar BHs are retained 
in GCs, or the MSP formation efficiency is much higher in the halo than in 
the disk.

We finally ran fully consistent simulations of the MSP population in the Galaxy.
We considered 30000 MSPs distributed in the disk plus several fast 
halo MSP populations predicted by our models. The disk and the halo 
populations were averaged over 1000  and 20 realizations respectively. 
We simulated observations with
SKA, taking into account for selection effects and limiting sensitivity
of the instrument. We found the distribution of projected 
distances from the galactic plane $h$ to be an effective
discriminant between the two populations. Fast halo MSPs, in fact, show a 
significant excess of objects at high $h$, in sharp contrast with
standard disk-born MSPs. This means that any excess
of MSPs at high $h$  would be a strong hint 
of the presence of a different MSP population.
Here we showed that MSP ejection from GC, due to dynamical interactions with IMBHs,
would easily produce such excess, and can therefore leave a clear 
imprint in future radio surveys. We quantified the statistical significance 
of halo MSP detection by performing tests on
synthetic samples drawn from the disk plus halo distribution. Results shown in
figure \ref{f8} and table \ref{tab2} are very encouraging. If we isolate the
distribution of MSPs observed at distances from the galactic plane 
larger than $h_{\rm cut}=3$ kpc, the halo population is, on average, detected 
at a 97\% confidence level for all models but L-I-${\cal A}$ and L-I-${\cal B}$. 
When $h_{\rm cut}=4$ kpc, detection confidence is boosted to better than 99\%
for all models. This is also true for the reduced SKA ${\cal B}$ configuration,
indicating that such detection will be easily feasible.
 

Our analysis is admittedly oversimplified in several respects. Most importantly,
the disk MSP population is hard to model, since $\sim100$ objects has been
detected to date. This introduces considerable uncertainties in our results. 
Note however that, unless disk MSPs are far more numerous than suggested
by present observation or extend far above the MW disk height, our basic
result, that an excess of MSPs located at large distances from the Galactic plane
would strongly support dynamical ejection due to IMBH residing in GCs, 
remains unaltered.

\section*{Acknowledgments}
We thank the anonimous referee for the valuable feedback that led to a 
significant improvement of the manuscript.
AS was supported by the DFG grant SFB/TR 7 "Gravitational Wave Astronomy".
NS acknowledges the support of ASI/INAF through grant I/009/10/0.
AP received support from the PRIN INAF 2010.
AS also acknowledges Piero Madau, Francesco Haardt and Monica Colpi 
for early discussions and suggestions that led to the start of this project.


\begin{thebibliography}{}

\item Archibald A. M. et al., 2009, Science, 324, 1411

\item Bacon D., Sigurdsson S. \& Davies M. B., 1996, MNRAS, 281, 830

\item Binney J. \& Tremaine S., "Galactic dynamics", Princeton, NJ, Princeton University Press, 1987, 747 p.

\item Bhattacharya D., \& van den Heuvel E.~P.~J., 1991, Physical Report, 203, 1 

\item Baumgardt H., Hut P., Makino J., McMillan S. \& Portegies Zwart S., 2003a, 
ApJ, 582, 21

\item Baumgardt H., Makino J., Hut P., McMillan S. \& Portegies Zwart S., 2003b, 
ApJ, 589, 25

\item Baumgardt H., Makino J. \& Hut P. 2005, ApJ, 620, 238

\item Beccari G., Pasquato M., De Marchi G., Dalessandro E., Trenti M. \& Gill M., 2010, ApJ, 713, 194

\item Bell E. F. et al., 2008, ApJ, 680, 295

\item Bromley B. C., Kenyon S. J., Geller M. J., Barcikowski E., Brown W. R. \& Kurtz M. J., 2006, ApJ, 653, 1194

\item Brown W. R., et. al. 2010, AJ, 139,59

\item Camilo F., Lorimer D. R., Freire P. C., Lyne A. G., Manchester R. N., 2000, ApJ, 535, 975

\item Camilo F \& Rasio F. A. 2005, ASPC, 328, 147

\item Chakrabarty D., 2006, AJ, 131, 2561

\item Cordes, J. M., \& Chernoff, D. F., 1997, ApJ, 482, 971

\item Cseh D., Kaaret P., Corbel S., Körding E., Coriat M., Tzioumis A. \& Lanzoni, B., 2010, MNRAS, 406, 1049

\item Devecchi, B., Colpi, M., Mapelli, M., \& Possenti, A. 2007,
  MNRAS, 380, 691

\item Faucher-Gigu\`ere C. A. \& Loeb A., 2010, JCAP, 1, 5

\item Gebhardt K., Rich R. M. \& Ho L. C., 2002, ApJ, 578, 41

\item Gebhardt K., Rich R. M. \& Ho L. C., 2005, ApJ, 634, 1093

\item Gillessen S., Eisenhauer F., Fritz T. K., Bartko H., Dodds-Eden K., Pfuhl O., Ott T. \& Genzel R., 2009, ApJ, 707, 114

\item Glebbeek E., Gaburov E., de Mink S. E., Pols O. R. \& Portegies Zwart S. F., 2009, A\&A, 497, 255

\item Gonzalez, M.~E., Stairs, I.~H., Ferdman, R.~D., et al.\ 2011, ApJ, 743, 102 

\item Harris W. E., 1996, AJ, 112, 1487

\item Hills J. G., 1975, AJ, 80, 809

\item Hills J. G., 1988, Nature, 331, 687

\item Ibata R. et al., 2009, ApJ, 699, 169

\item Ivanova N., Heinke C. O., Rasio F. A., Belczynski K. \& Fregeau, J. M., 2008a, MNRAS, 386, 553

\item Ivanova N., Heinke C. O. \& Rasio F. A., 2008b, IAUS, 246, 316

\item King A. R., Davies M. B. \& Beer M. E., 2003, MNRAS, 345, 678

\item Kobayashi S., Hainick Y., Sari R. \& Rossi E. M., 2012, ApJ, 748, 105

\item Kramer M., Xilouris K. M., Lorimer D. R., Doroshenko O., Jessner A., Wielebinski R., Wolszczan A. \& Camilo F., 1998, ApJ, 501, 270

\item Lazio J., 2009, arXiv:0910:0632

\item Lorimer D. R., 2005, LRR, 8, 7

\item Lorimer D. R. et al., 2006, MNRAS, 372, 777

\item Lorimer D. R., 2008, LRR, 11, 8

\item Lyne A. G. et al., 1998, MNRAS, 295, 743

\item Maccarone, T.~J., \& Servillat, M. 2008, MNRAS, 389, 379

\item McConnell D., Deshpande A. A., Connors T. \& Ables J. G., 2004, MNRAS, 348, 1409

\item McLaughlin D. E., 2000, ApJ, 539, 618

\item McLaughlin D. E., 2003, Proceedings of the ESO Workshop Held in Garching, Germany, 27-30 August 2002, ESO ASTROPHYSICS SYMPOSIA. ISBN 3-540-40472-4. Edited by M. Kissler-Patig. Springer-Verlag, 2003, p. 329

\item McLaughlin D. E. \& Fall S. M., 2008, ApJ, 679, 1272 

\item Miller M. C. \& Hamilton D. P., 2002, MNRAS, 330, 232 

\item Muno M. P. et al., 2003, ApJ, 589, 225

\item Noyola E., Gebhardt K. \& Bergmann M., 2008, ApJ, 676, 1008 

\item Nucita A. A., de Paolis F., Ingrosso G., Carpano S. \& Guainazzi M., 2008, A\&A, 478, 763

\item O'Leary R. M., Rasio F. A., Fregeau J. M., Ivanova N. \& O'Shaughnessy R., 2006, ApJ, 637, 937

\item Pasquato M., Trenti M., De Marchi G., Gill M., Hamilton D. P., Miller M. C., Stiavelli M. \& van der Marel R. P., 2009, ApJ, 699, 1511

\item Pooley, D., et al. 2003, ApJL, 591, L131

\item Portegies Zwart S. \& McMillan S. L. W., 2002, ApJ, 576, 899

\item Portegies Zwart S., Baumgardt H., Hut P., Makino J. \& McMillan S. L. W., 2004, Nature, 428, 724

\item Portegies Zwart S., 2005, n "Joint Evolution of Black Holes and Galaxies" of the Series in High Energy Physics, Cosmology and Gravitation. IOP Publishing, Bristol and Philadelphia, 2005, eds M. Colpi, V.Gorini, F.Haardt and U.Moschella

\item Sari R., Kobayashi S. \& Rossi E. M., 2010, ApJ, 708, 605

\item Sartore N., et al. 2010, A\&A, 510, 23

\item Sesana A., Haardt F. \& Madau P., 2007, MNRAS, 379, 45

\item Smits R., et al. 2009, A\&A, 490, 1161

\item Smits R., Tingay S. J., Wex N., Kramer M. \& Stappers B., 2011, A\&A, 528, 108

\item Sollima A., Bellazzini M., Smart R. L., Correnti M., Pancino E., Ferraro F. R. \& Romano D., 2009, MNRAS, 396, 2183

\item Story, S.~A., Gonthier, P.~L., \& Harding, A.~K.\ 2007, ApJ, 671, 713 

\item Tremaine S. et al., 2002, ApJ, 574, 740

\item Trenti M., 2008, IAUS, 246, 256

\item Umbreit S., Fregeau J. M. \& Rasio F. A., 2009, arXiv:0910.5293

\item van den Bosch R., de Zeeuw T., Gebhardt K., Noyola E. \& van de Ven G., 2006, ApJ, 641, 852

\item van den Heuvel E.~P.~J., 1995, Journal of Astrophysics and Astronomy, 16, 255 


\item van den Heuvel E. P. J. \& van Paradijs J., 1988, Nature, 334, 227

\item van der Marel R. P. \& Anderson J., 2010, ApJ, 710, 1063

\end{thebibliography}
\end{document}